\documentclass{aastex}
\usepackage{textcomp}
\usepackage{rotating}
\usepackage{epstopdf}
\title{Measurement of boron and carbon fluxes in cosmic rays with the PAMELA experiment.}
\author
{O. Adriani$^{1,2}$, G. C. Barbarino$^{3,4}$, G. A. Bazilevskaya$^{5}$, R. Bellotti$^{6,7}$, M. Boezio$^{8}$, \\
E. A. Bogomolov$^{9}$, M. Bongi$^{1,2}$, V. Bonvicini$^{8}$, S. Bottai$^{2}$, A. Bruno$^{6,7}$, F. Cafagna$^{7}$,\\
D. Campana$^{4}$, R. Carbone$^{8}$, P. Carlson$^{13}$, M. Casolino$^{10,14}$, G. Castellini$^{15}$, \\
I. A. Danilchenko$^{12}$, C. De Donato$^{10,11}$, C. De Santis$^{10,11}$, N. De Simone$^{10}$, V. Di Felice$^{10,19}$, \\
V. Formato$^{8,16}$, A. M. Galper$^{12}$, A. V. Karelin$^{12}$, S. V. Koldashov$^{12}$, S. Koldobskiy$^{12}$, \\
S. Y. Krutkov$^{9}$, A. N. Kvashnin$^{5}$, A. Leonov$^{12}$, V. Malakhov$^{12}$, L. Marcelli$^{10,11}$, \\
M. Martucci$^{11,20}$, A. G. Mayorov$^{12}$, W. Menn$^{17}$, M. Merg\'{e}$^{10,11}$, V. V. Mikhailov$^{12}$, \\
E. Mocchiutti$^{8}$, A. Monaco$^{6,7}$,  N. Mori$^{2,21}$, R. Munini$^{8,16}$, G. Osteria$^{4}$, F. Palma$^{10,11}$, \\ 
B. Panico$^{4}$, P. Papini$^{2}$, M. Pearce$^{13}$, P. Picozza$^{10,11}$, C. Pizzolotto$^{18,19\, *}$,  M. Ricci$^{20}$, \\
S. B. Ricciarini$^{2,15}$, L. Rossetto$^{13}$, R. Sarkar$^{22\, *}$, V. Scotti$^{3,4}$, M. Simon$^{17}$,  R. Sparvoli$^{10,11}$,\\
P. Spillantini$^{1,2}$, Y. I. Stozhkov$^{5}$, A. Vacchi$^{8}$, E. Vannuccini$^{2}$, G. I. Vasilyev$^{9}$,\\ 
S. A. Voronov$^{12}$, Y. T. Yurkin$^{12}$, G. Zampa$^{8}$, N. Zampa$^{8}$, V. G. Zverev$^{12}$ \\
}
\affil{$^{1}$University of Florence, Department of Physics and Astronomy, I-50019 Sesto Fiorentino, Florence, Italy}
\affil{$^{2}$INFN, Sezione di Florence, I-50019 Sesto Fiorentino, Florence, Italy}
\affil{$^{3}$University of Naples ``Federico II'', Department of Physics, I-80126 Naples, Italy}
\affil{$^{4}$INFN, Sezione di Naples,  I-80126 Naples, Italy}
\affil{$^{5}$Lebedev Physical Institute, RU-119991, Moscow, Russia}
\affil{$^{6}$University of Bari, Department of Physics, I-70126 Bari, Italy}
\affil{$^{7}$INFN, Sezione di Bari, I-70126 Bari, Italy}
\affil{$^{8}$INFN, Sezione di Trieste, I-34149 Trieste, Italy}
\affil{$^{9}$Ioffe Physical Technical Institute,  RU-194021 St. Petersburg, Russia}
\affil{$^{10}$INFN, Sezione di Rome ``Tor Vergata'', I-00133 Rome, Italy}
\affil{$^{11}$University of Rome ``Tor Vergata'', Department of Physics,  I-00133 Rome, Italy}
\affil{$^{12}$National Research Nuclear University MEPhI, RU-115409 Moscow}
\affil{$^{13}$KTH, Department of Physics, and the Oskar Klein Centre for Cosmoparticle Physics, AlbaNova University Centre, SE-10691 Stockholm, Sweden}
\affil{$^{14}$RIKEN, Advanced Science Institute, Wako-shi, Saitama, Japan}
\affil{$^{15}$IFAC, I-50019 Sesto Fiorentino, Florence, Italy}
\affil{$^{16}$University of Trieste, Department of Physics, I-34147 Trieste, Italy}
\affil{$^{17}$Universit\"{a}t Siegen, Department of Physics, D-57068 Siegen, Germany}
\affil{$^{18}$INFN, Sezione di Perugia, I-06123 Perugia, Italy}
\affil{$^{19}$Agenzia Spaziale Italiana (ASI) Science Data Center, Via del Politecnico snc I-00133 Rome, Italy}
\affil{$^{20}$INFN, Laboratori Nazionali di Frascati, Via Enrico Fermi 40, I-00044 Frascati, Italy}
\affil{$^{21}$Centro Siciliano di Fisica Nucleare e Struttura della Materia (CSFNSM), Viale A. Doria 6, I-95125 Catania, Italy}
\affil{$^{22}$Indian Centre for Space Physics, 43, Chalantika, Garia Station Road, Kolkata– 700 084, West Bengal, India}
\affil{$^{*}$Previously at INFN, Sezione di Trieste, I-34149 Trieste, Italy}

\begin{abstract}
The propagation of cosmic rays inside our galaxy plays a fundamental role in shaping their injection spectra into those
observed at Earth. One of the best tools to investigate this issue is the ratio of fluxes for secondary and primary species.
The boron-to-carbon (B/C) ratio, in particular, is a sensitive probe to investigate propagation mechanisms. This paper 
presents new measurements of the absolute fluxes of boron and carbon nuclei, as well as the B/C ratio, from the PAMELA space
experiment. The results span the range 0.44 - 129 GeV/n in kinetic energy for data taken in the period July 2006 - March 2008.
\end{abstract}

\begin{document}
\section{Introduction}
Propagation in the interstellar medium (ISM) significantly affects the spectrum of galactic cosmic rays. After being accelerated by high-energy
astrophysical processes such as supernovae explosions, cosmic rays are injected into the interstellar space, propagate through it 
and eventually reach the Earth where they are detected. The multitude of physical processes that cosmic rays undergo
during propagation (e.g. diffusion, spallation, emission of synchrotron radiation etc.) shape the injection spectra and chemical
composition into the observed values. A detailed knowledge of these processes is therefore needed in order to interpret the
experimental data in terms of source parameters, or in estimating the expected background when searching for contributions from new 
sources.

There is still a relatively high degree of uncertainty regarding the physical processes relevant to propagation of cosmic rays and the impact 
of experimental uncertainties on the determination of propagation parameters (see \cite{ref:uncertainties} and references therein). The propagation is usually modelled
in terms of a diffusive transport equation \cite{ref:propagation}. The equation contains terms which account for diffusion in the irregular
galactic magnetic field, convection due to the galactic wind, energy losses, re-acceleration (modelled as diffusion in momentum space), 
spallation and radioactive decay, and source terms. Some parameters of the equation are simply related to directly 
measurable quantities unrelated to cosmic rays, and thus they can be obtained from independent measurements (e.g. the density of atomic 
hydrogen in the ISM, which is needed in order to estimate the spallation rate, can be measured by means of 21 cm radio surveys). Other 
parameters are obtained by fitting distributions derived from numerical propagation models like GALPROP \cite{ref:galprop1, ref:galprop2} or DRAGON  \cite{ref:dragon}
to direct cosmic ray measurements.

In order to test and tune the propagation models, a particularly useful measurable quantity is the secondary to primary flux ratio. Primary nuclei
are those accelerated by cosmic ray sources such as supernova remnants, whereas secondaries are those produced in interactions of primaries with the ISM
during propagation. The boron to carbon flux ratio (B/C) has been widely studied. Since boron is produced in negligible quantities by stellar 
nucleosynthesis processes \cite{ref:nucleosynthesis}, almost all of the observed boron is believed to be from spallation 
reactions of CNO primaries on atomic and molecular H and He present in the ISM. The B/C flux ratio is therefore a clean and direct probe of 
propagation mechanisms, and it is considered as the ``standard tool'' for studying propagation models \cite{ref:propreview, ref:bcleakybox}.

The B/C flux ratio, as well as the absolute boron and carbon fluxes, have been measured by balloon-borne \cite{ref:freier, ref:atic, ref:cream, ref:tracer} and by
space-based experiments \cite{ref:heao-3-c2, ref:crn, ref:voyager, ref:ams01, ref:acecris, ref:ams02}, with different techniques and spanning various
energy ranges from about 80 MeV/n up to a few TeV/n. Even if the spread in the measurements and their associated errors makes it difficult
to clearly discriminate between the various models or to tightly constrain model parameters, there is a general
consensus about several points. The relative abundance of the light elements Li, Be and B in cosmic rays is
significantly higher than in the solar system \cite{ref:propagacecris}. This supports the idea of creation by spallation
reactions in ISM. The B/C flux ratio has a peak value at $\sim 1$ GeV/n, which can favour a model with distributed stochastic re-acceleration 
\cite{ref:reaccelleakybox}. The B/C flux ratio decreases at high energies and its shape, in diffusive models, is mainly determined by 
the energy dependence of the diffusion coefficient \cite{ref:diffcoeff}.

In this paper, a new set of measurements of boron and carbon fluxes as well as the B/C flux ratio obtained with the PAMELA instrument in the
kinetic energy range 0.44 - 129 GeV/n during the solar minimum period spanning from July 2006 to March 2008 are presented. The study of solar 
modulation effects on the low-energy component of the spectra over a longer time interval will be the subject of a future publication. After a brief 
description of the PAMELA detector system, the analysis techniques and an evaluation of systematic uncertainties are presented, followed by a discussion of the results.

\section{The PAMELA detector}
\begin{figure}[htbp]
\begin{center}
\includegraphics[width=7.5cm]{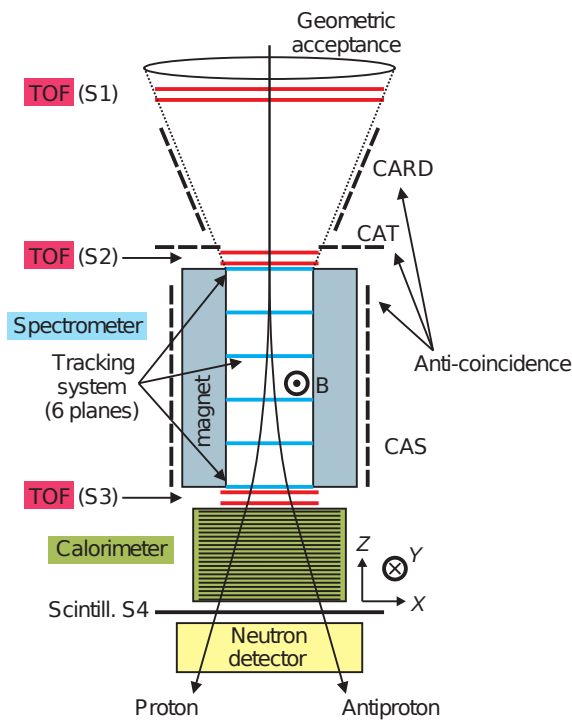}
\end{center}
\vspace{-0.5cm}
\caption{Schematic view of the PAMELA apparatus.}
\label{fig:PAMELA}
\end{figure}

A schematic view of the PAMELA detector system \cite{ref:PAMELA} is shown in Figure \ref{fig:PAMELA}. The design was chosen to meet the main scientific goal of precisely 
measuring the light components of the cosmic ray spectrum in the energy range starting from tens of MeV up to 1 TeV (depending on particle species), 
with a particular focus on antimatter.
To this end, the design is optimized for $|Z| = 1$ particles and to provide a high lepton-hadron discrimination power.
The core of the instrument is a magnetic spectrometer \cite{ref:tracker} made by six double-sided silicon microstrip tracking layers placed in the 
bore of a permanent magnet. The read-out pitch of the silicon sensors is 51 $\mu$m in the X (bending) view and 66.5 $\mu$m
in the Y view. The spectrometer provides information about the magnetic rigidity $\rho = pc/(Ze)$ of the particle (where $p$ and $Z$ are 
the particle momentum and the electric charge, respectively). Six layers of plastic scintillator paddles 
arranged in three X-Y planes (S1, S2 and S3 in Figure \ref{fig:PAMELA}) placed above and below the magnetic cavity constitute the 
Time-Of-Flight (TOF) system \cite{ref:tof, ref:tofelectronics}. The flight time of particles is measured with a time resolution of 250 ps for $|Z| = 1$ 
particles and about 70 ps for boron and carbon nuclei \cite{ref:toftestbeam}. This allows albedo particles to be rejected and, in combination 
with the track length information obtained from the tracking system, precise measurement of the particle velocity, $\beta=v/c$. The TOF 
scintillators can identify the absolute particle charge up to oxygen by means of six independent ionisation measurements.
The tracking system and the upper TOF system are shielded by an anticoincidence system (AC) \cite{ref:ac} made of plastic scintillators and arranged in 
three sections (CARD, CAT and CAS in Figure \ref{fig:PAMELA}), which allows spurious triggers generated by secondary particles to be rejected during offline 
data analysis.
A sampling electromagnetic calorimeter \cite{ref:calo1, ref:calo2} is placed below S3. It consists of 22 modules, each comprising a tungsten converter 
layer placed between two layers equipped with single-sided silicon strip detectors with orthogonal read-out strips. The total depth of the calorimeter is $16.3 X_0$,
while the readout pitch of the strips is 2.44 mm. The calorimeter measures the energy of electrons and positrons, and gives a 
lepton/hadron rejection power of $\sim 10^5$ by means of topological shower analysis, thanks to its fine lateral and longitudinal segmentation. 
A tail-catcher scintillating detector (S4) and a neutron detector placed below the calorimeter help to further improve the rejection power.

The geometric factor of the apparatus, defined by the magnetic cavity, is energy dependent because of the track curvature induced by the magnetic field, and increases 
as the energy of the particle increases. However, for rigidities above 1 GV it varies only by a few per mil, reaching the value of 21.6 cm$^2$ sr at the highest rigidity.

The PAMELA apparatus was launched on June 15th 2006, and has been continuously taking data since then. It is hosted as a piggyback payload 
on the Russian satellite Resurs-DK1, which executes a 70\textdegree \, semi-polar orbit. The orbit was elliptical with variable height between 
350 and 620 km up to 2010, after which it was converted to the current circular orbit with height about 600 km.

\section{Data analysis}

\subsection{Data processing}
\label{sect:dataprocessing}
The event reconstruction routines require silicon strips to be gathered into clusters. A ``seed'' 
strip is defined as a strip with a signal to noise ratio (S/N) greater than 7; it is grouped with its neighbouring ``signal'' strips with S/N $>$ 4 to form a cluster. For each 
cluster an estimate of the particle impact point is obtained by means of an analog position finding algorithm \cite{ref:tracker}. The original reconstruction routines 
were conceived and tuned to deal with $|Z| \sim 1$ particles. However the higher ionisation energy losses of boron and carbon in the silicon layers of the tracking system 
saturate the front-end electronics, leading to a degradation of the performance of position finding with respect to $|Z| \sim 1$ particles. 
A different position finding algorithm has thus been implemented for this saturation regime. For each cluster of silicon strips, the saturated strips have 
been treated as if read-out system was digital, and the impact point has been evaluated as the geometric centre. The associated spatial resolution
can be approximated as the readout pitch over $\sqrt{12}$, which translates to $\sim$ 14 $\mu$m for the X (bending) view and $\sim$ 19 $\mu$m for the Y view.
The associated MDR (Maximum Detectable Rigidity\footnote{The MDR is defined as the rigidity with an associated 100\% error due to the finite spatial resolution of the
spectrometer}) is $\sim 250$ GV.

Prior to event reconstruction, the clusters with an associated energy release less than 5 MIP\footnote{1 MIP is defined 
here as the most probable energy release of a $|Z|=1$ minimum ionising particle} have been removed. This helps to eliminate clusters associated with 
delta rays and light secondary particles, e.g. backscattered particles from the calorimeter. There is a twofold effect: the tracking efficiency is increased
since the tracking algorithm has less clusters to deal with, and the energy dependence of the tracking efficiency is reduced at high
energies by removing backscattering clusters, which are mainly produced by high energy primaries interacting in the calorimeter.

\subsection{Event selection}
\label{sect:selections}
In order to be able to reliably measure the magnetic rigidity, events with a single track in the spectrometer containing at least 4 hits in the X view and 
3 hits in the Y view have been selected. A good $\chi^2$ value for the fitted track was required. The $\chi^2$ distribution is energy dependent and thus the 
selection criterion has been calibrated in order to obtain a constant efficiency of about 90\% over the whole energy range, in particular at low energies where 
multiple scattering leads to generally higher $\chi^2$ values. Reconstructed tracks were required
to lie entirely inside a fiducial volume with bounding surfaces 0.15 cm from the magnet walls. Galactic events were selected by imposing that the lower edge 
of the rigidity bin to which the event belongs exceeds the critical rigidity, $\rho_c$, defined as 1.3 times the cutoff rigidity $\rho_{SVC}$ computed in the 
St\"{o}rmer vertical approximation \cite{ref:stoermer} as $\rho_{SVC} = 14.9/L^2$, where $L$ is the McIlwain $L$-shell parameter \cite{ref:Lshell} obtained by 
using the Resurs-DK1 orbital information and the IGRF magnetic field model \cite{ref:igrf}. The South Atlantic Anomaly region has been included in the analysis.
Reconstructed particle trajectories were required to be down-going according to the TOF. No selections on the hit pattern in the TOF paddles or AC were made, 
since this can lead to very low efficiencies due to the production of delta rays in the aluminum dome of the pressurized vessel in which PAMELA is hosted. 
This introduces a contamination from secondaries produced in hadronic interactions of primaries in the dome. This effect has been accounted for using 
Monte Carlo simulations.

\begin{figure}[htbp]
\begin{center}
\includegraphics[width=8cm]{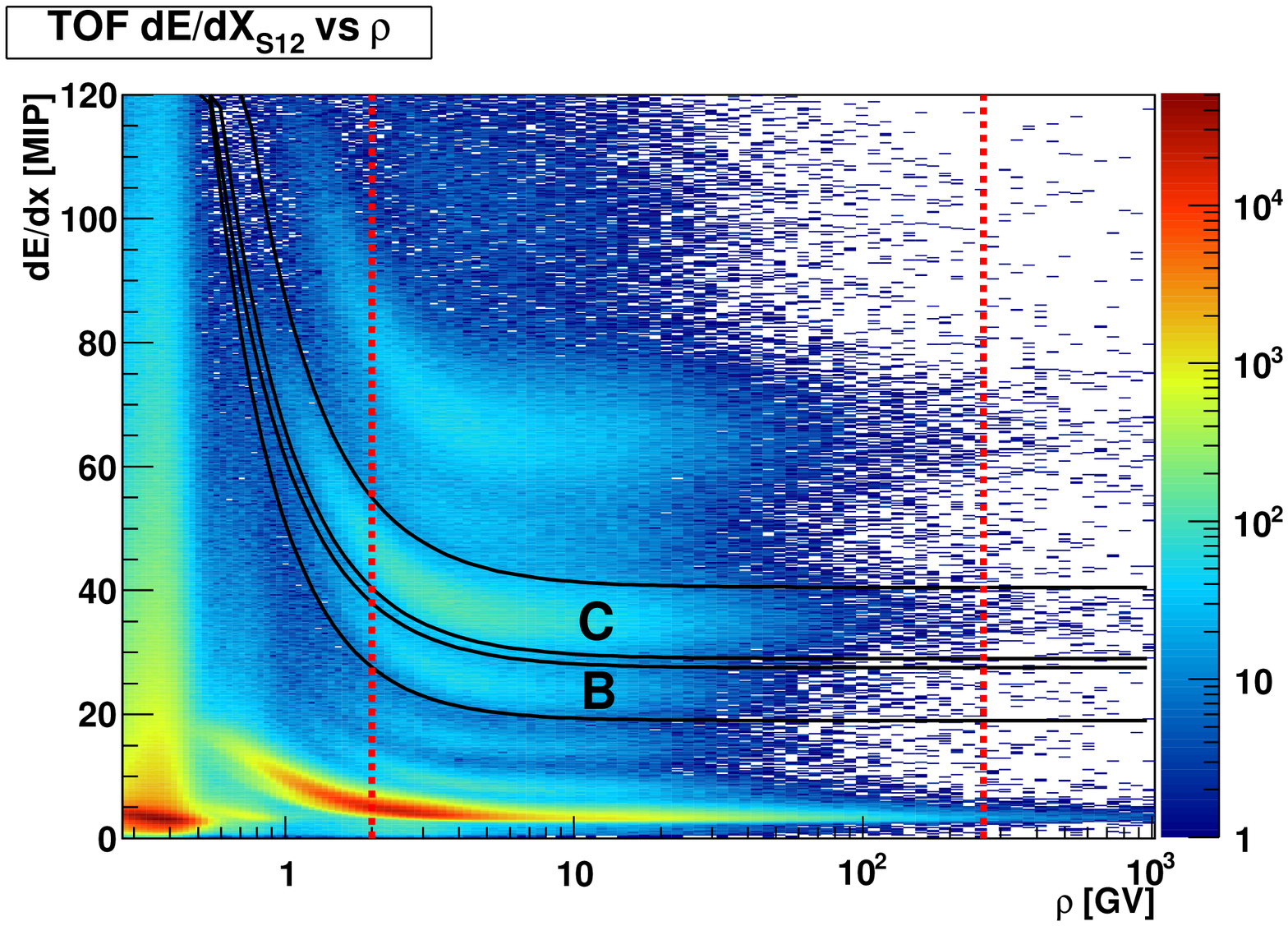}
\includegraphics[width=8cm]{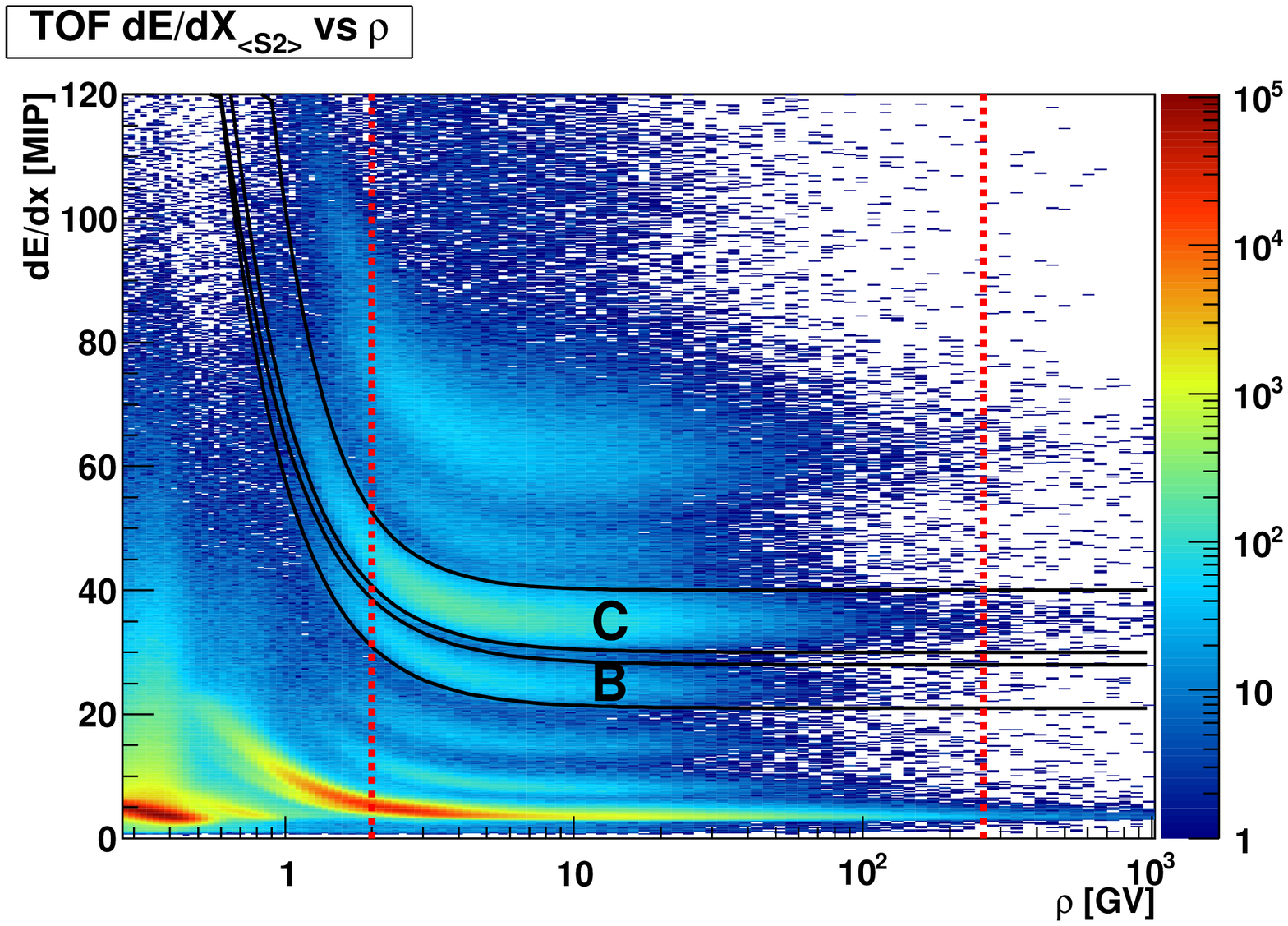}
\includegraphics[width=8cm]{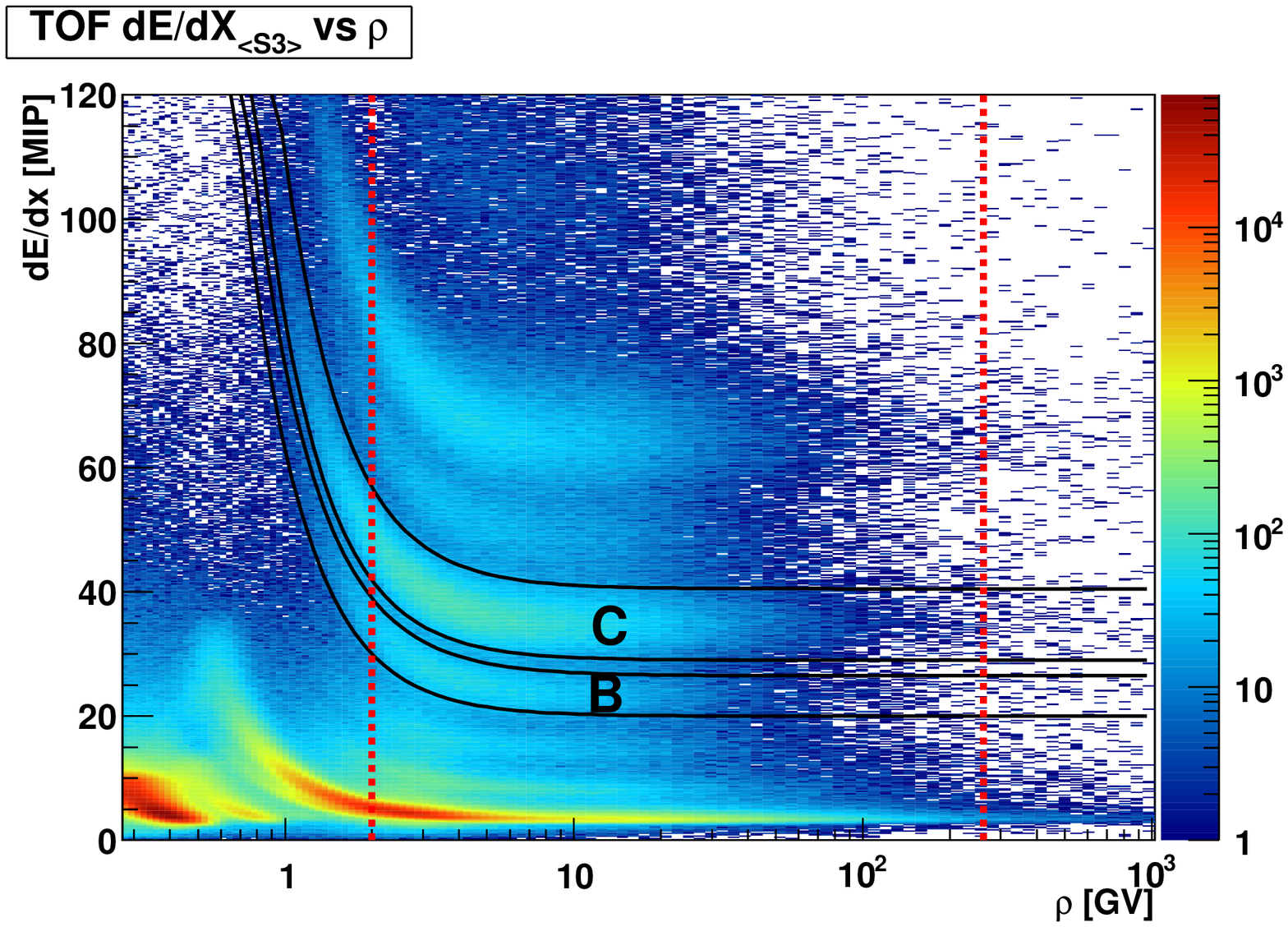}
\end{center}
\vspace{-0.5cm}
\caption{Charge selection bands for S12, $\langle$S2$\rangle$ and $\langle$S3$\rangle$ as a function of rigidity. The red vertical dotted 
lines denote the upper and lower rigidity limits of this analysis. The absence of relativistic protons in this sample is due to the 
5 MIP cluster selection described in Section \ref{sect:dataprocessing}.}
\label{fig:chargesel}
\end{figure}

Boron and carbon events have been selected by means of ionisation energy losses in the TOF system. Charge consistency has been required between 
S12\footnote{S12 is the lowest of the two layers constituting S1; the upper layer S11 was used for efficiency measurement as explained in Section \ref{sect:efficiencies}}
and $\langle$S2$\rangle$ and $\langle$S3$\rangle$ (the arithmetic mean of the ionisations for the two layers constituting S2 and S3, respectively).
Requiring charge consistency above and below the tracking system rejected events interacting in the silicon layers. The selection bands as functions 
of the rigidity measured by the spectrometer are shown in Figure \ref{fig:chargesel}.

In order to assess the presence of possible contamination in the selected samples, the above selection cuts have been applied to boron and
carbon samples independently selected by means of S11 (the upper layer of S1) and the first silicon layer of the calorimeter. The probabilities of 
misidentifying a carbon nucleus as boron and vice versa are about $3 \cdot 10^{-4}$ and $10^{-3}$, respectively, over the whole energy range considered
in this analysis. 
Stricter analysis criteria were imposed by narrowing the selection bands. When properly corrected by the selection efficiency (see Section 
\ref{sect:efficiencies}), the event counts showed no statistically significant deviation from that obtained using the standard selection. The
contamination is therefore assumed to be negligible.
Selected events have been binned according to the rigidity measured by the magnetic spectrometer.

\subsection{Efficiencies}
\label{sect:efficiencies}
The tracking efficiency has been evaluated with flight data and Monte Carlo simulations using a methodology similar to that described in 
\cite{ref:solarmodulation}. Two samples of boron and carbon were selected by means of a $\beta$ dependent requirement on ionisation energy losses
in the TOF system. Fiducial containment was verified using calorimeter information. Firstly, non-interacting events penetrating deeply into the calorimeter
were identified, and a straight track fitted. Then, the rigidity of the nucleus was derived from the $\beta$ measured by the TOF and used to 
back-propagate the track through the spectrometer magnetic field up to the top of the apparatus. The containment criteria were applied to this 
back-extrapolated track. The tracking efficiency was determined for this sample of non-interacting nuclei as a function of the rigidity derived from 
$\beta$. The 70 ps resolution of the TOF system for carbon leads to $\Delta \beta / \beta \sim 2\%$ at $\beta=0.9$ \cite{ref:toftestbeam}. Bin 
folding effects on the efficiency have therefore been neglected.

Due to the calorimeter selection criteria described above, the efficiency is measured for a non-isotropically distributed sample,
while the fluxes impinging on PAMELA are isotropic. Moreover, a possible energy dependence of the efficiency at relativistic energies cannot be
accounted for by an efficiency measured as a function of $\beta$. To account for these effects, a simulation of the PAMELA apparatus based on 
GEANT4 \cite{ref:geant4-1, ref:geant4-2} has been used to estimate the isotropic, rigidity-dependent tracking efficiency which is subsequently 
divided by a Monte Carlo efficiency obtained using the same procedure as the experimental efficiency. 
The resulting ratio, which has an almost constant value of about 0.97, has been used as the correction factor for the experimental efficiency. The 
constancy of the ratio results from an isotropic efficiency that is also almost constant above 10 GV because of the data processing procedures
described in Section \ref{sect:dataprocessing}.

The efficiencies for the selection of down-going particles and for charge selection have been estimated using flight data exclusively. The down-going 
requirement is 100\% efficient due to the 70 ps resolution of the TOF system. To evaluate the charge selection efficiency, the redundancy
of the PAMELA subdetectors has been exploited.
Two samples of boron and carbon have been tagged requiring charge consistency on S11 and on the first silicon layer of the calorimeter. These two detectors 
are placed at the two extrema of the apparatus, so this selection rejects interactions which change the reconstructed charge of the incident particle.
The resulting efficiencies have a peak value of $\sim 75\%$ at 3 GV and then decrease at high energies towards an almost constant value
of about 50\% for boron and 60\% for carbon above some tens of GV.

\begin{figure}[htbp]
\begin{center}
\includegraphics[width=15cm]{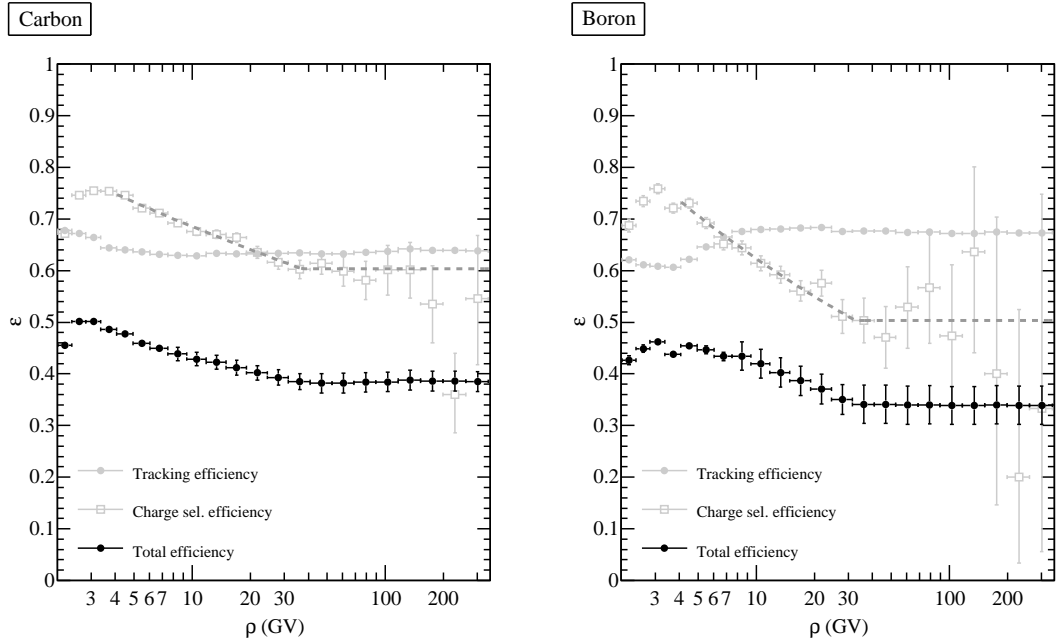}
\end{center}
\vspace{-0.5cm}
\caption{Selection efficiencies as functions of rigidity. The dashed line is a fit of the charge selection efficiency above 3 GV with a power law at low 
rigidities and a constant value at high rigidities. The slope, the break point and the normalization are free parameters of the fit. The fitted charge
selection efficiency is used to compute the total efficiency above 3 GeV/n (about 7.6 GV for C and $^{10}$B and 8.4 GV for $^{11}$B) in order to smooth
the statistical fluctuations.}
\label{fig:chargeseleff}
\end{figure}

The tracking and the charge selection efficiencies are shown in Figure \ref{fig:chargeseleff} together with the total selection efficiency.

The measurement of the charge selection efficiency sets the lower rigidity limit for fluxes to 2 GV, corresponding to about 0.44 GeV/n for $^{10}$B
and $^{12}$C. Below this threshold charge confusion in the calorimeter selection becomes too large to be able to reliably tag
pure boron and carbon samples for an efficiency measurement.

The effects of a possible contamination in the efficiency samples tagged with S11 and the calorimeter (S11+CALO tag) have been investigated by 
considering a single TOF layer and measuring the charge selection efficiency both on the event set tagged with S11+CALO and on a purer sample
obtained by adding the other TOF planes to the S11+CALO tag. The two efficiencies were found to be consistent within statistical errors for
each layer. No effect due to contamination in the S11+CALO tagged set was observed.

\subsection{Corrections}
\label{sect:corrections}
The selected boron and carbon samples are contaminated by secondaries produced during fragmentation processes occurring in the aluminum dome on top 
of the pressurized vessel hosting PAMELA. This effect has been studied with a Monte Carlo calculation based on the FLUKA code
\cite{ref:fluka} by simulating the cosmic spectra for C and O, which are the main contributors to the contamination. The resulting contamination
is of the order of $10^{-3}$ for carbon, whereas for boron it ranges from about 5\% at some GV up to about 20\% at $\sim$ 200 GV, coming
mainly from spallation of carbon.

After subtracting the contamination, the rigidity distributions of boron and carbon events have been corrected for folding effects using
a Bayesian procedure \cite{ref:unfolding}, in order to obtain the distributions at the top of payload. These effects include possible
rigidity displacements at high energies due to the finite position resolution of the silicon tracking layers and the energy loss of low-energy 
nuclei traversing the apparatus. The smearing matrix was derived using the GEANT4 simulations.

\begin{figure}[htbp]
\begin{center}
\includegraphics[width=15cm]{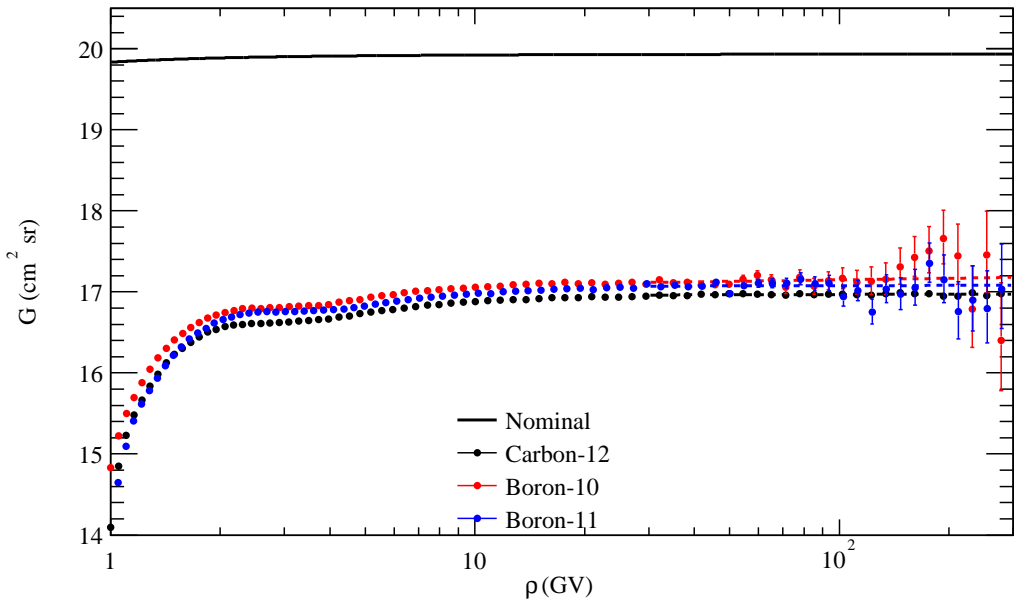}
\end{center}
\vspace{-0.5cm}
\caption{Effective geometrical factors including the fiducial containment criterion and the correction for interactions of primary particles above the tracker.
The dashed lines are fits used to obtain asymptotic values at high energy.}
\label{fig:geomfact}
\end{figure}

Interactions with the aluminum dome also remove primaries from the selected samples. Elastic scattering processes can remove primaries from the instrument 
acceptance or slow them down so that they are swept out by the magnetic field; inelastic scattering can destroy the primary. A correction factor for these
effects has been evaluated using the FLUKA simulations, and applied to the unfolded event count. The correction is almost flat above 10 GV and 
amounts to 15\% for carbon and 14\% for boron, increasing at lower energies because of energy loss. These numbers have been treated as corrections
to the geometrical factor for the two nuclear species. The resulting geometrical factors are shown in Figure \ref{fig:geomfact}.

Energy loss in the apparatus may lower the measured rigidity below the critical rigidity, leading to rejection of galactic nuclei with
initial rigidity above the critical one. A ``cutoff correction factor'' for each nuclear species was computed by assigning 
a random cutoff value (distributed as observed for in-flight values) to events simulated with GEANT4 and deriving the fraction of rejected events.
This correction factor rises from about 0.97 at 2 GV to unity (i.e. no correction) at 3 GV and above.

\subsection{Live time}
The live time of the apparatus is measured by on-board clocks and has been evaluated as a function of the vertical cutoff as the time spent 
in regions where the critical rigidity is below the lower limit of the rigidity bin. The total live time is constant at a value of $\sim 3.14 \times 10^7$ s
for rigidities above 20 GV and decreases at lower rigidities because of the shorter time spent by the satellite in high latitude (i.e. low cutoff)
regions down to $\sim 1.00 \times 10^7$ s at 2 GV. The overall error on live time determination is less than 0.2\%, and has therefore been neglected.

\subsection{Geometrical factor}
Due to the requirement of track containment inside a fiducial volume (see Section \ref{sect:selections}), the effective geometrical factor
turns out to be lower than the nominal one, and assumes a constant value of 19.9 cm$^2$ sr above 1 GV. This value has been
cross-checked using two different numerical methods. The first one is a numerical computation of the integral defining the geometrical factor 
\cite{ref:sullivan}, taking into account the curvature of the track due to the magnetic field, while the second method relies on a Monte Carlo simulation 
\cite{ref:sullivan}. The two methods yield results differing by less than 0.1\%. This error has also been neglected.

\subsection{Flux computation}
The fluxes have been computed both as functions of rigidity and as functions of kinetic energy per nucleon. For each bin $i$, the event count $\Delta N'_i$ 
corrected for the effects described in Section \ref{sect:corrections} was divided by the live time $\Delta T_i$, the effective geometrical factor $\tilde G_i$, the total 
selection efficiency $\epsilon_i$ and the bin width $\Delta \rho_i$ or $\Delta E_i$. The flux expressed as a function of rigidity is computed as:
\begin{displaymath}
 \phi(\rho_i) = \frac{\Delta N'_i}{\Delta T_i \, \tilde G_i \, \epsilon_i \, \Delta \rho_i} \, ,
\end{displaymath}
while as a function of kinetic energy per nucleon:
\begin{displaymath}
 \phi(E_i) = \frac{\Delta N'_i}{\Delta T_i \, \tilde G_i \, \epsilon_i \, \Delta E_i} \, .
\end{displaymath}
For boron, the latter formula needs to properly account for isotopic composition, as explained in the next section.

\subsection{Isotopic composition for boron}
\label{sect:boronisotopes}
In cosmic rays, both the isotopes $^{10}$B and $^{11}$B are present in comparable quantities. Since the event selection did not 
distinguish between them and since the events are binned according to their rigidity, a given value for the isotopic composition of boron must be 
assumed in order to perform the measurements as functions of kinetic energy per nucleon. Large uncertainties plague the available estimates of the
isotopic composition of boron. Direct measurements are available only at relatively low energies \cite{ref:smili,ref:isomax,ref:ams01isotopes}. 
Galactic propagation models predict a high-energy value for the $^{10}$B fraction (i.e., $^{10}$B/($^{10}$B + $^{11}$B)) which is weakly dependent 
on kinetic energy per nucleon and whose consensus value is $\tilde{F_B} = 0.35 \pm 0.15$. This value has been used in this analysis for the whole
energy range.

The boron flux has been evaluated considering two different hypotheses: pure $^{10}$B and pure $^{11}$B. Assuming a binning in kinetic energy per nucleon,
the corresponding binning in rigidity for each of the two hypotheses has been derived. Event selection, efficiency measurements, flux computation
and corrections have then been performed in the same way for the two binnings. The two boron fluxes are combined to obtain the final flux, considering that
each bin of each flux distribution contains $^{10}$B and $^{11}$B events with the same rigidity but different energy due to the different masses.
Consequently, in each bin the isotopic fraction does not resemble the usual fraction expressed as a function of kinetic energy per 
nucleon, and a simple bin-by-bin linear combination of the two fluxes using $\tilde{F_B}$ as the weight would lead to an incorrect result. A fraction 
$F_B(\rho)$ has been derived by means of Monte Carlo simulations and used as a weight in order to linearly combine the two boron fluxes bin by bin
and obtain a final flux. A detailed description of the calculation is presented in Appendix \ref{app:boronisotopes}.

\section{Results}
\begin{figure}[htbp]
\begin{center}
\includegraphics[width=13cm]{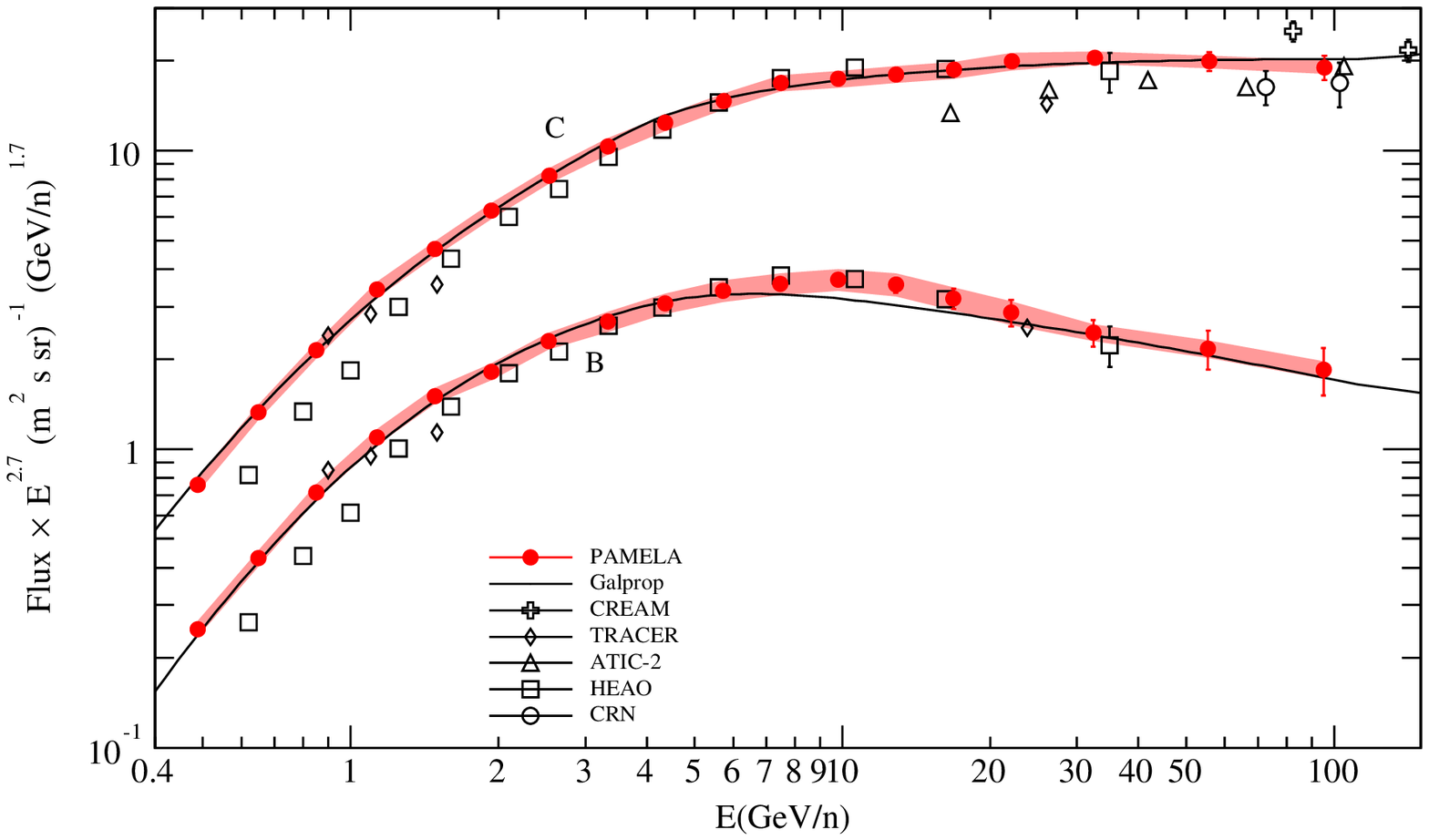}
\includegraphics[width=13cm]{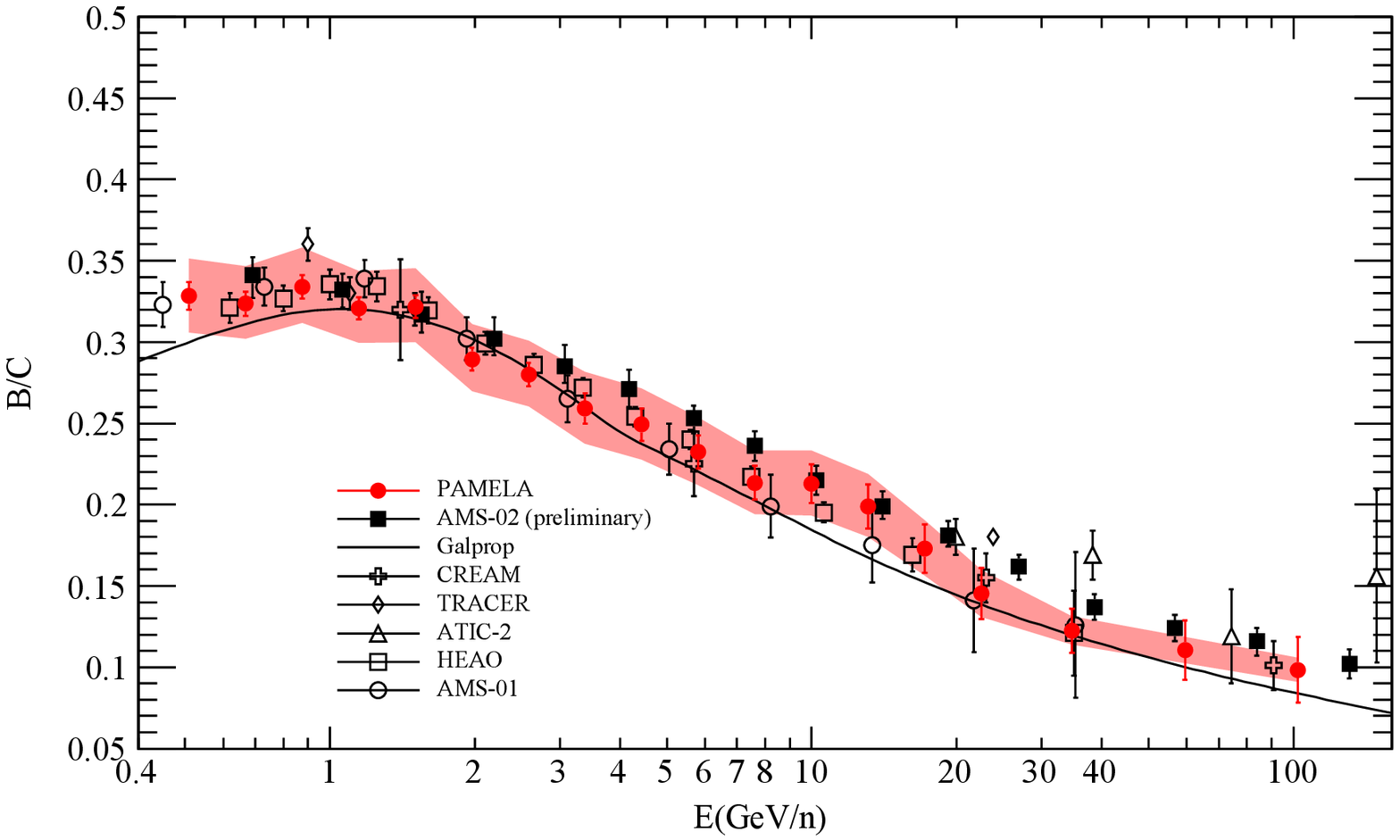}
\end{center}
\vspace{-0.5cm}
\caption{Absolute boron and carbon fluxes multiplied by E$^{2.7}$ (upper panel) and B/C flux ratio (lower panel) as measured by PAMELA, together 
with results from other experiments (AMS02 \cite{ref:ams02}, CREAM \cite{ref:cream}, TRACER \cite{ref:tracer}, ATIC-2 \cite{ref:atic}, HEAO \cite{ref:heao-3-c2},
AMS01 \cite{ref:ams01}, CRN \cite{ref:crn}) and a theoretical calculation based on GALPROP (see Section \ref{sect:discussion}), as functions of kinetic energy 
per nucleon. For PAMELA data the error bars represent the statistical error and the shaded area is the overall systematic uncertainty summarized in 
Appendix \ref{app:syserrs}.}
\label{fig:fluxes}
\end{figure}

\begin{figure}[htbp]
\begin{center}
\includegraphics[width=13cm]{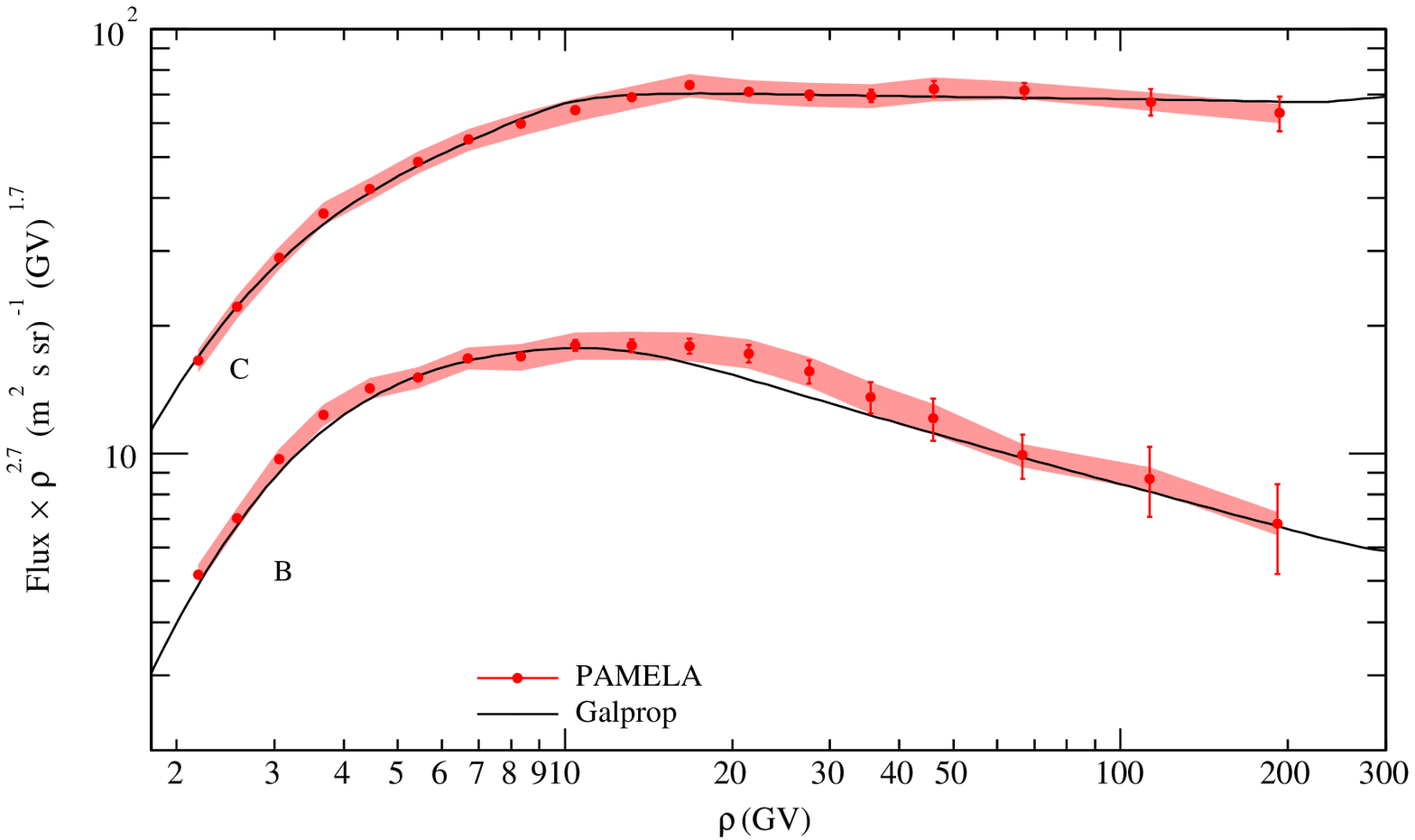}
\includegraphics[width=13cm]{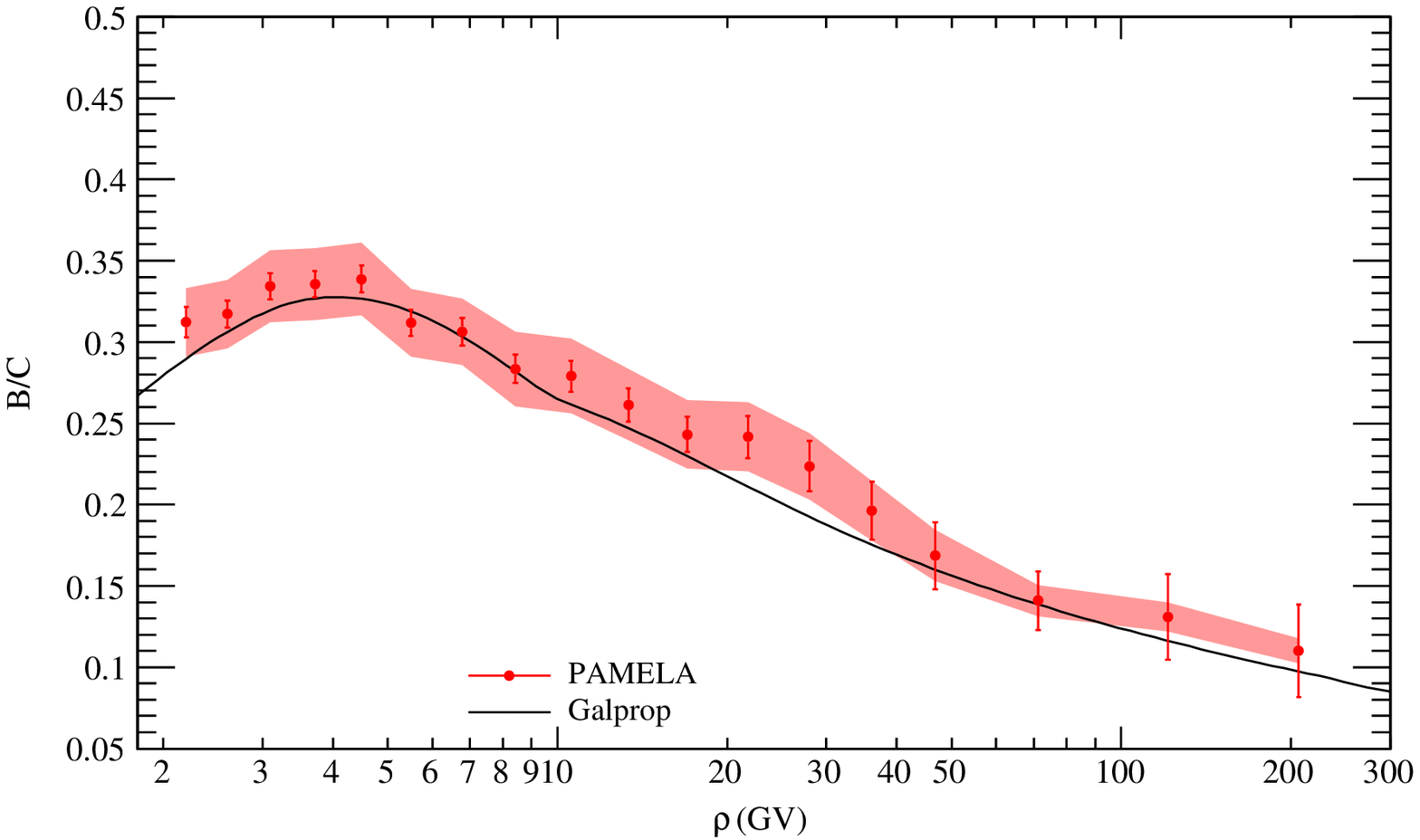}
\end{center}
\vspace{-0.5cm}
\caption{Absolute boron and carbon fluxes multiplied by $\rho^{2.7}$ (upper panel) and B/C flux ratio (lower panel) as measured by PAMELA, together 
with a theoretical calculation based on GALPROP (see Section \ref{sect:discussion}), as functions of rigidity. The error bars represent the statistical error
and the shaded area is the overall systematic uncertainty summarized in Appendix \ref{app:syserrs}, except for the boron mixing error which does not affect the rigidity-dependent 
boron flux.}
\label{fig:fluxesrig}
\end{figure}

The observed number of selected boron and carbon events, the absolute fluxes and the B/C flux ratio are reported in Tables \ref{table:fluxes} and 
\ref{table:fluxesrig}. The quoted systematic uncertainties are discussed in detail in Appendix \ref{app:syserrs}. The fluxes and the B/C ratio are 
also shown in Figures \ref{fig:fluxes} and \ref{fig:fluxesrig} along with measurements from other experiments and a theoretical calculation 
based on GALPROP. The details of the calculation are described in Section \ref{sect:discussion}. The mean kinetic energy $<E>$ and the mean rigidity 
$<\rho>$ for each bin have been computed according to \cite{ref:bincenters} using an iterative procedure starting from the middle point of each bin.
The resulting mean energies and rigidities for boron and carbon differ by less than 1\%, and have been considered to be equal.

The discrepancies with other experiments at low energies can be reasonably ascribed to solar modulation effects. The data used for this 
analysis were taken by PAMELA during an unusually quiet solar minimum period, resulting in an enhanced flux of galactic cosmic rays at 
low energies in the heliosphere, which has already been observed for protons \cite{ref:solarmodulation} and nuclei 
\cite{ref:solarmodulationnucleicris}. Above 6 GeV/n the fluxes are in overall agreement with the other available measurements, especially
with those from HEAO and CREAM. A power-law fit above 20 GeV/n results in a spectral index $\gamma_B=3.01 \pm 0.13$ for boron and $
\gamma_C=2.72 \pm 0.06$ for carbon.

\begin{sidewaystable}[htbp]
\resizebox{\textwidth}{!}{ 
\begin{tabular}{| c | c | c | c | c| c | c | c |}
\hline
\hline
 Kinetic energy & $\langle E \rangle$ & C events & C flux & $^{10}$B events & $^{11}$B events & B flux & B/C \\
 at top of payload &  &  & value $\pm$ stat. $\pm$ syst. & & & value $\pm$ stat. $\pm$ syst. & value $\pm$ stat. $\pm$ syst. \\
 (GeV/n) & (GeV/n) & & (GeV/n m$^2$ s sr)$^{-1}$ & & & (GeV/n m$^2$ s sr)$^{-1}$ & \\
\hline
0.44 - 0.58 & 0.49 & 5146 &$(5.26 \pm 0.08 \pm 0.26)$              & 1566  & 1795& $(1.73 \pm 0.04^{+ 0.09}_{- 0.08})$ & $(3.28 \pm 0.09^{+ 0.23}_{- 0.22}) \cdot 10^{-1}$ \\ 
0.58 - 0.76 & 0.65 & 6651 &$(4.27 \pm 0.05 \pm 0.21)$              & 1955  & 2092& $(1.38 \pm 0.03^{+ 0.07}_{- 0.06})$ & $(3.24 \pm 0.07^{+ 0.23}_{- 0.21}) \cdot 10^{-1}$ \\ 
0.76 - 1.00 & 0.85 & 7359 &$(3.30 \pm 0.04 \pm 0.16)$              & 2300  & 2320& $(1.102 \pm 0.020^{+ 0.059}_{- 0.050})$ & $(3.34 \pm 0.07^{+ 0.24}_{- 0.22}) \cdot 10^{-1}$ \\ 
1.00 - 1.30 & 1.13 & 7578 &$(2.45 \pm 0.03 \pm 0.12)$              & 2351  & 2248& $(7.85 \pm 0.14^{+ 0.42}_{- 0.36}) \cdot 10^{-1}$ & $(3.21 \pm 0.07^{+ 0.23}_{- 0.21}) \cdot 10^{-1}$ \\ 
1.30 - 1.71 & 1.50 & 7033 &$(1.612 \pm 0.019 \pm 0.078)$           & 2281  & 2166& $(5.18 \pm 0.10^{+ 0.29}_{- 0.24}) \cdot 10^{-1}$ & $(3.22 \pm 0.07^{+ 0.24}_{- 0.22}) \cdot 10^{-1}$ \\ 
1.71 - 2.24 & 1.94 & 6369 &$(1.057 \pm 0.013 \pm 0.051)$           & 1960  & 1737& $(3.06 \pm 0.06^{+ 0.17}_{- 0.15}) \cdot 10^{-1}$ & $(2.89 \pm 0.07^{+ 0.22}_{- 0.20}) \cdot 10^{-1}$ \\ 
2.24 - 2.93 & 2.53 & 5673 &$(6.70 \pm 0.09 \pm 0.32) \cdot 10^{-1}$& 1691  & 1553& $(1.88 \pm 0.04^{+ 0.11}_{- 0.09}) \cdot 10^{-1}$ & $(2.80 \pm 0.07^{+ 0.21}_{- 0.20}) \cdot 10^{-1}$ \\ 
2.93 - 3.84 & 3.34 & 4795 &$(3.99 \pm 0.06 \pm 0.20) \cdot 10^{-1}$& 1350  & 1202& $(1.03 \pm 0.03 \pm 0.07) \cdot 10^{-1}$ & $(2.59 \pm 0.09^{+ 0.23}_{- 0.21}) \cdot 10^{-1}$ \\ 
3.84 - 5.03 & 4.36 & 3990 &$(2.32 \pm 0.04 \pm 0.12) \cdot 10^{-1}$& 1078  & 945 & $(5.8 \pm 0.2 \pm 0.4) \cdot 10^{-2}$ & $(2.49 \pm 0.10^{+ 0.22}_{- 0.21}) \cdot 10^{-1}$ \\ 
5.03 - 6.60 & 5.73 & 3270 &$(1.31 \pm 0.02 \pm 0.07) \cdot 10^{-1}$& 811   & 704 & $(3.05 \pm 0.12^{+ 0.23}_{- 0.22}) \cdot 10^{-2}$ & $(2.32 \pm 0.10^{+ 0.21}_{- 0.20}) \cdot 10^{-1}$ \\ 
6.60 - 8.65 & 7.49 & 2717 &$(7.32 \pm 0.14 \pm 0.38) \cdot 10^{-2}$& 612   & 540 & $(1.56 \pm 0.07 \pm 0.12) \cdot 10^{-2}$ & $(2.134 \pm 0.10^{+ 0.20}_{- 0.19}) \cdot 10^{-1}$ \\ 
8.65 - 11.3 & 9.81 & 2048 &$(3.65 \pm 0.08 \pm 0.19) \cdot 10^{-2}$& 454   & 369 & $(7.8 \pm 0.4 \pm 0.6) \cdot 10^{-3}$ & $(2.128 \pm 0.12 \pm 0.20) \cdot 10^{-1}$ \\ 
11.3 - 14.9 & 12.9 & 1337 &$(1.81 \pm 0.05 \pm 0.10) \cdot 10^{-2}$& 253   & 217 & $(3.6 \pm 0.2 \pm 0.3) \cdot 10^{-3}$ & $(1.99 \pm 0.13^{+ 0.20}_{- 0.19}) \cdot 10^{-1}$ \\ 
14.9 - 19.5 & 16.9 & 851 &$(9.0 \pm 0.3 \pm 0.5) \cdot 10^{-3}$    & 149   & 121 & $(1.56 \pm 0.12^{+ 0.13}_{- 0.12}) \cdot 10^{-3}$ & $(1.73 \pm 0.15 \pm 0.17) \cdot 10^{-1}$ \\ 
19.5 - 25.5 & 22.1 & 571 &$(4.6 \pm 0.2 \pm 0.3) \cdot 10^{-3}$    & 85    & 69  & $(6.7 \pm 0.7 \pm 0.6) \cdot 10^{-4}$ & $(1.45 \pm 0.16 \pm 0.14) \cdot 10^{-1}$ \\ 
25.5 - 43.8 & 32.6 & 590 &$(1.67 \pm 0.07 \pm 0.07) \cdot 10^{-3}$ & 79    & 65  & $(2.1 \pm 0.2 \pm 0.1) \cdot 10^{-4}$ & $(1.22 \pm 0.13 \pm 0.09) \cdot 10^{-1}$ \\ 
43.8 - 75.3 & 55.7 & 225 &$(3.8 \pm 0.3 \pm 0.2) \cdot 10^{-4}$    & 31    & 24  & $(4.2 \pm 0.6 \pm 0.3) \cdot 10^{-5}$ & $(1.11 \pm 0.18 \pm 0.08) \cdot 10^{-1}$ \\ 
75.3 - 129 & 95.6  & 86 &$(8.5 \pm 0.8 \pm 0.4) \cdot 10^{-5}$     & 9     & 7   & $(8.4 \pm 1.5 \pm 0.5) \cdot 10^{-6}$ & $(10 \pm 2 \pm 0.7) \cdot 10^{-2}$ \\ 
\hline
\end{tabular}
}
\caption{Observed number of events, absolute fluxes and the B/C flux ratio as function of kinetic energy per nucleon. Both the event counts for pure $^{10}$B and pure $^{11}$B hypotheses are reported.}
\label{table:fluxes}
\end{sidewaystable}

\begin{sidewaystable}[htbp]
\resizebox{\textwidth}{!}{ 
\begin{tabular}{| c | c | c | c | c | c | c |}
\hline
\hline
 Rigidity & $\langle \rho \rangle$ & C events & C flux & B events & B flux & B/C \\
 at top of payload &  &  & value $\pm$ stat. $\pm$ syst. &  & value $\pm$ stat. $\pm$ syst. & value $\pm$ stat. $\pm$ syst. \\
 (GV) & (GV) &  & (GV m$^2$ s sr)$^{-1}$ &  & (GV m$^2$ s sr)$^{-1}$ & \\
\hline
2.02 - 2.38 & 2.19 & 5146 &$(2.01 \pm 0.03 \pm 0.10)$                  & 1566 & $(6.26 \pm 0.16 \pm 0.29) \cdot 10^{-1}$ & $(3.12 \pm 0.09 \pm 0.21) \cdot 10^{-1}$ \\ 
2.38 - 2.82 & 2.57 & 6651 &$(1.73 \pm 0.02 \pm 0.08)$                  & 1955 & $(5.49 \pm 0.13 \pm 0.25) \cdot 10^{-1}$ & $(3.17 \pm 0.08 \pm 0.21) \cdot 10^{-1}$ \\ 
2.82 - 3.37 & 3.06 & 7359 &$(1.413 \pm 0.017 \pm 0.068)$               & 2300 & $(4.72 \pm 0.10 \pm 0.21) \cdot 10^{-1}$ & $(3.34 \pm 0.08 \pm 0.22) \cdot 10^{-1}$ \\ 
3.37 - 4.06 & 3.67 & 7578 &$(1.093 \pm 0.013 \pm 0.053)$               & 2351 & $(3.67 \pm 0.08 \pm 0.17) \cdot 10^{-1}$ & $(3.35 \pm 0.08 \pm 0.22) \cdot 10^{-1}$ \\ 
4.06 - 4.93 & 4.45 & 7033 &$(7.44 \pm 0.09 \pm 0.36) \cdot 10^{-1}$    & 2281 & $(2.52 \pm 0.05 \pm 0.11) \cdot 10^{-1}$ & $(3.39 \pm 0.08 \pm 0.22) \cdot 10^{-1}$ \\ 
4.93 - 6.06 & 5.44 & 6369 &$(5.00 \pm 0.06 \pm 0.24) \cdot 10^{-1}$    & 1960 & $(1.56 \pm 0.04 \pm 0.07) \cdot 10^{-1}$ & $(3.12 \pm 0.08 \pm 0.21) \cdot 10^{-1}$ \\ 
6.06 - 7.50 & 6.70 & 5673 &$(3.23 \pm 0.04 \pm 0.16) \cdot 10^{-1}$    & 1691 & $(9.9 \pm 0.2 \pm 0.5) \cdot 10^{-2}$ & $(3.06 \pm 0.08 \pm 0.21) \cdot 10^{-1}$ \\ 
7.50 - 9.36 & 8.34 & 4795 &$(1.95 \pm 0.03 \pm 0.10) \cdot 10^{-1}$    & 1350 & $(5.52 \pm 0.15 \pm 0.34) \cdot 10^{-2}$ & $(2.83 \pm 0.09 \pm 0.23) \cdot 10^{-1}$ \\ 
9.36 - 11.8 & 10.4 & 3990 &$(1.143 \pm 0.018 \pm 0.058) \cdot 10^{-1}$ & 1078 & $(3.19 \pm 0.10 \pm 0.20) \cdot 10^{-2}$ & $(2.79 \pm 0.10 \pm 0.23) \cdot 10^{-1}$ \\ 
11.8 - 15.0 & 13.2 & 3270 &$(6.49 \pm 0.11 \pm 0.33) \cdot 10^{-2}$    & 811  & $(1.70 \pm 0.06 \pm 0.11) \cdot 10^{-2}$ & $(2.61 \pm 0.10 \pm 0.22) \cdot 10^{-1}$ \\ 
15.0 - 19.1 & 16.8 & 2717 &$(3.64 \pm 0.07 \pm 0.19) \cdot 10^{-2}$    & 612  & $(8.8 \pm 0.4 \pm 0.6) \cdot 10^{-3}$ & $(2.43 \pm 0.11 \pm 0.21) \cdot 10^{-1}$ \\ 
19.1 - 24.5 & 21.4 & 2048 &$(1.82 \pm 0.04 \pm 0.10) \cdot 10^{-2}$    & 454  & $(4.4 \pm 0.2 \pm 0.3) \cdot 10^{-3}$ & $(2.42 \pm 0.13 \pm 0.21) \cdot 10^{-1}$ \\ 
24.5 - 31.5 & 27.6 & 1337 &$(9.1 \pm 0.3 \pm 0.5) \cdot 10^{-3}$       & 253  & $(2.02 \pm 0.13 \pm 0.15) \cdot 10^{-3}$ & $(2.24 \pm 0.15 \pm 0.20) \cdot 10^{-1}$ \\ 
31.5 - 40.8 & 35.6 & 851  &$(4.51 \pm 0.16 \pm 0.24) \cdot 10^{-3}$    & 149  & $(8.9 \pm 0.7 \pm 0.7) \cdot 10^{-4}$ & $(1.96 \pm 0.18 \pm 0.18) \cdot 10^{-1}$ \\ 
40.8 - 52.9 & 46.1 & 571  &$(2.32 \pm 0.10 \pm 0.13) \cdot 10^{-3}$    & 85   & $(3.9 \pm 0.4 \pm 0.3) \cdot 10^{-4}$ & $(1.7 \pm 0.2 \pm 0.16) \cdot 10^{-1}$ \\ 
52.9 - 89.5 & 67.1 & 590  &$(8.4 \pm 0.4 \pm 0.3) \cdot 10^{-4}$       & 79   & $(1.18 \pm 0.14 \pm 0.07) \cdot 10^{-4}$ & $(1.41 \pm 0.18 \pm 0.10) \cdot 10^{-1}$ \\ 
89.5 - 152 & 113   & 225  &$(1.92 \pm 0.14 \pm 0.08) \cdot 10^{-4}$    & 31   & $(2.5 \pm 0.5 \pm 0.1) \cdot 10^{-5}$ & $(1.3 \pm 0.3 \pm 0.09) \cdot 10^{-1}$ \\ 
152 - 260 & 193    & 86   &$(4.3 \pm 0.4 \pm 0.2) \cdot 10^{-5}$       & 9    & $(4.7 \pm 1.1 \pm 0.3) \cdot 10^{-6}$ & $(1.1 \pm 0.3 \pm 0.08) \cdot 10^{-1}$ \\ 
\hline
\end{tabular}
}
\caption{Observed number of events, absolute fluxes and the B/C flux ratio as function of rigidity.}
\label{table:fluxesrig}
\end{sidewaystable}

\section{Discussion}
\label{sect:discussion}
A comprehensive and detailed study of the results presented above is beyond the scope of this paper. The following discussion is intentionally limited to
a single propagation model in order to compute an estimate of the most significant propagation parameters from the PAMELA boron and carbon data. Results 
may vary when considering different models or propagation software packages.

\begin{figure}[htbp]
\begin{center}
\includegraphics[width=7.5cm]{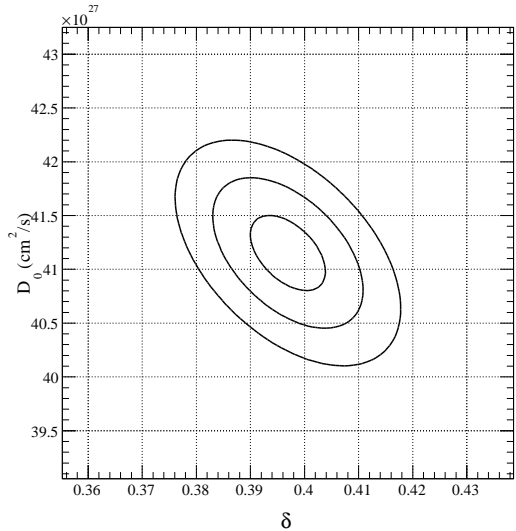}
\end{center}
\vspace{-0.5cm}
\caption{Contour plot of the 1-, 2- and 3-sigma confidence levels for $\delta$ and $D_0$.}
\label{fig:contourplot}
\end{figure}

The data presented in the previous section as a function of kinetic energy per nucleon has been fitted with a diffusive cosmic ray propagation model using
the GALPROP code interfaced with the MIGRAD minimizer in the MINUIT2 minimization package distributed within the ROOT framework \cite{ref:root}. Only 
a few parameters have been left free because of the high computation time required for multiple GALPROP runs. The values for the other parameters have
been taken from \cite{ref:galprop2}. The diffusion coefficient is found to have a fitted slope value of $\delta = 0.397 \pm 0.007$ and a normalization factor 
$D_0 = (4.12 \pm 0.04) \cdot 10^{28}$ cm$^2$/s. Other fitted parameters are the solar modulation parameter in the force-field approximation $\Phi=(0.40 \pm 0.01)
$ GV and the overall normalization of the fluxes $N=1.04\pm 0.03$. The result of the fit is shown in Figures \ref{fig:fluxes} and \ref{fig:fluxesrig}. A 
contour plot of the confidence intervals for $\delta$ and $D_0$ is shown in Figure \ref{fig:contourplot}.

The fitted value for $\delta$ falls between the predicted values for Kolmogorov ($\delta=1/3$) and Kraichnan
($\delta=1/2$) diffusion types, thus the PAMELA data cannot distinguish between these two types.

\section{Acknowledgements}
We acknowledge support from The Italian Space Agency (ASI), Deutsches Zentrum f\"{u}r Luft- und Raumfahrt (DLR), 
The Swedish National Space Board, The Swedish Research Council, The Russian Space Agency (Roscosmos) and The Russian 
Science Foundation. 

\appendix
\section{Systematic uncertainties}
\label{app:syserrs}
\begin{figure}[htbp]
\begin{center}
\includegraphics[width=8cm]{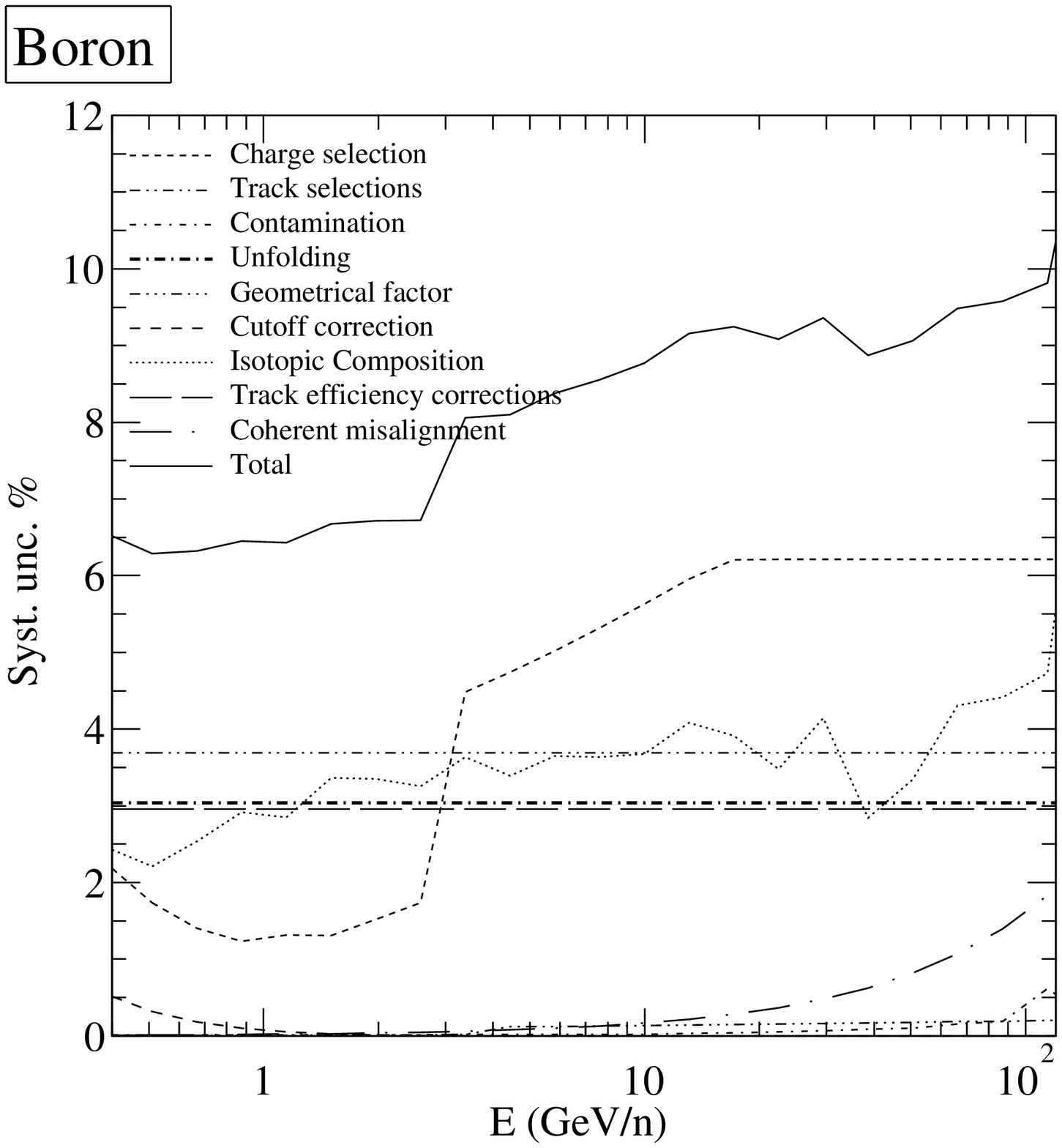}
\includegraphics[width=8cm]{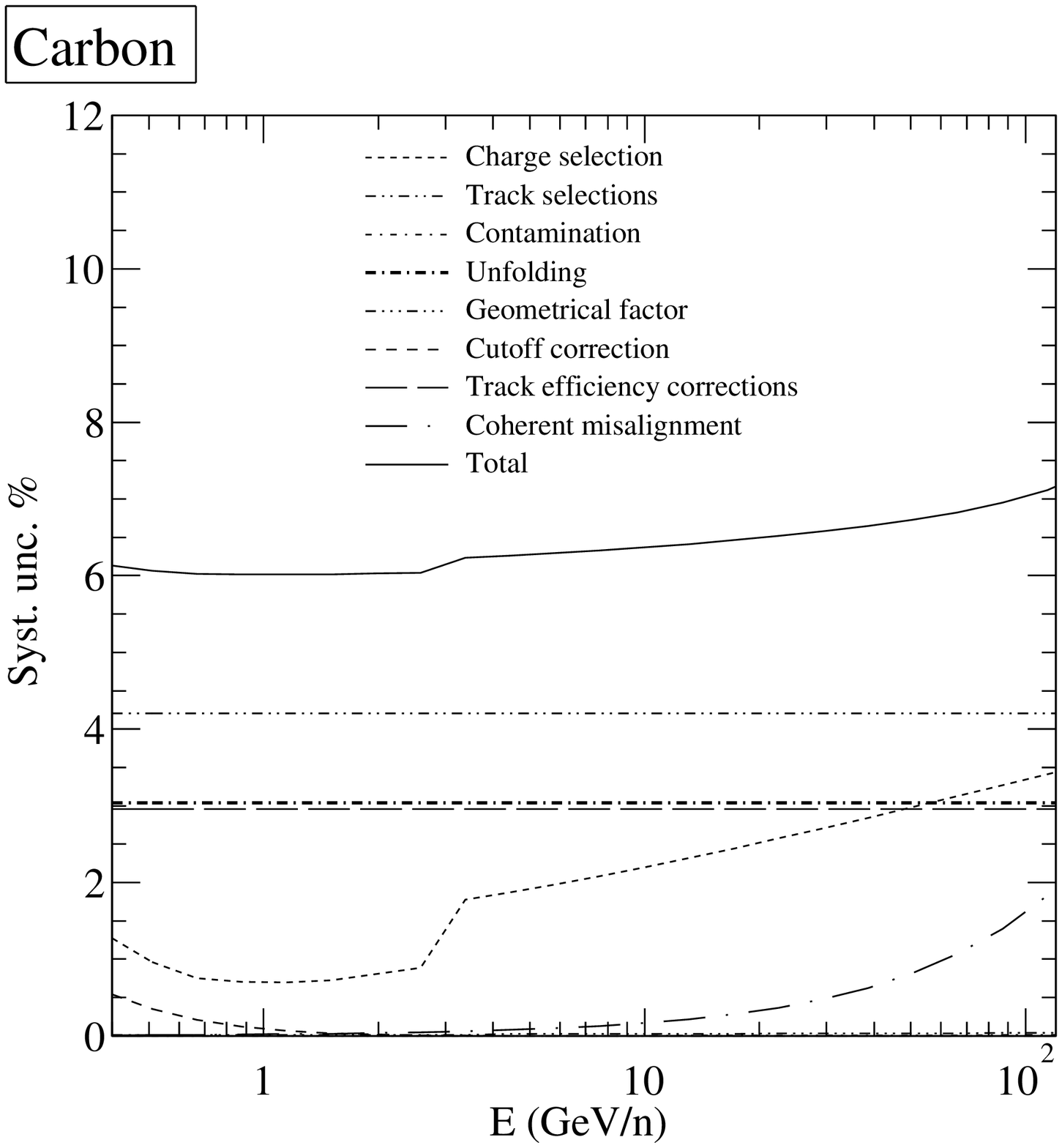}
\end{center}
\vspace{-0.5cm}
\caption{Systematic uncertainties for absolute fluxes. The total contribution is computed as the quadratic sum of the individual terms. The track 
selections term is the quadratic sum of the contributions from statistics and from fiducial containment. The contributions
of the track efficiency correction and of the unfolding have been slightly shifted apart from their 3\% value for the sake of readability.}
\label{fig:syserr}
\end{figure}

The following contributions to the systematic uncertainty have been considered:
\begin{itemize}
 \item {\it Selection efficiencies}: the measurement of the tracking and charge selection efficiencies from flight data is performed using samples
       of finite size. The associated statistical error has been propagated to the flux as a systematic uncertainty.
 \item {\it Fiducial containment}: the finite tracking resolution of the calorimeter can lead to a contamination of the tracking efficiency sample
       by events coming from outside the fiducial acceptance, and possibly also crossing the magnet walls. These can in principle be eliminated by 
       further restricting the fiducial volume for both event selection and efficiency measurement, but this would significantly reduce the 
       sample sizes. The chosen approach is to use protons from both flight and simulated data to measure the tracking efficiency 
       for both the fiducial volume defined in Section \ref{sect:selections} and a more restrictive one. Their relative difference is taken as an 
       estimate of the systematic uncertainty, which is about 2\%. Monte Carlo simulations give results for boron and carbon which are consistent
       with the one obtained with protons. The uncertainty is propagated to the final flux.
 \item {\it Monte Carlo correction factor for the tracking efficiency}: this correction factor should introduce only relatively small errors,
       since it is computed as the ratio of two Monte Carlo efficiencies. Systematic effects should largely cancel out. The correction factor is
       constant at 0.97 for both boron and carbon. That this factor remains constant at high rigidity is due to the isotropic efficiency being constant 
       at relativistic rigidities. A conservative factor of 3\% has been taken as an estimate of the systematic uncertainty on the flux because of this 
       correction factor.
 \item {\it Residual coherent misalignment of the spectrometer}: the spectrometer alignment procedure results in a residual coherent misalignment
       producing a systematic shift in the measured rigidity. The error estimation procedure is described in the Supporting Online Material of 
       \cite{ref:protonsscience} . This error has been propagated to the measured flux. It is negligible at low energy and increases up to about 
       2\% at 250 GV.
 \item {\it Cutoff, contamination and geometrical factor corrections}: all these factors have been evaluated on finite-size samples, so they are
       affected by a statistical error which has been propagated to the flux as a systematic uncertainty.
 \item {\it Unfolding}: the unfolding error has been assessed by means of the procedure described in the Supporting Online Material of 
       \cite{ref:protonsscience}, comparing a given initial spectrum and an unfolded Monte Carlo simulation. The two were found to be in agreement 
       within 3\%, so this value has been taken as the unfolding contribution to the flux error.
 \item {\it Isotopic composition of boron}: the uncertainty associated with this poorly known parameter has been propagated to the flux by 
       assuming the extreme values of 0.2 and 0.5 for the $^{10}$B fraction and taking the difference between these fluxes and the one obtained
       with $\tilde{F_B}=0.35$ as the estimated upper and lower errors on the flux. This error affects only the measurement expressed as a function
       of kinetic energy per nucleon.
\end{itemize}
The overall uncertainty has been estimated as the quadratic sum of the above terms in the hypothesis of uncorrelated errors. A summary plot is
shown in Figure \ref{fig:syserr}.

\section{Isotopic composition of boron}
\label{app:boronisotopes}
In this analysis the events have been binned according to their rigidity as measured by the magnetic spectrometer. Given that the
event selection does not distinguish between the two isotopes $^{10}$B and $^{11}$B, each bin is populated by $^{10}$B 
and $^{11}$B events with approximately the same rigidity (within the bin limits) but different kinetic energy per nucleon
because of the different mass numbers. Consequently, the isotopic composition in a given bin is not described by the $^{10}$B
fraction $F_B$ expressed as a function of kinetic energy per nucleon $E$:
\begin{equation}
  F_B(E) = \frac{\phi^{^{10}B}(E)}{\phi^{^{10}B}(E) + \phi^{^{11}B}(E)}\, ,
\end{equation}
where $\phi^{^{10}B}(E)$ and $\phi^{^{11}B}(E)$ are the fluxes of $^{10}$B and $^{11}$B respectively. A fraction expressed
as a function of rigidity must then be derived in order to correctly account for the isotopic composition in each bin:
\begin{equation}
  F_B(\rho) = \frac{\phi^{^{10}B}(\rho)}{\phi^{^{10}B}(\rho) + \phi^{^{11}B}(\rho)}\, .
\end{equation}
Using rigidity bins of finite size leads to:
\begin{equation}
  F_B(\rho_i) = \frac{\Delta N^{^{10}B}(\rho_i)}{\Delta N^{^{10}B}(\rho_i)+\Delta N^{^{11}B}(\rho_i)}\, ,
  \label{eq:rigfracbins}
\end{equation}
where $F_B(\rho_i)$ is the $^{10}$B fraction for the $i$-th rigidity bin centered at $\rho_i$, while $\Delta N^{^{10}B}(\rho_i)$
and $\Delta N^{^{11}B}(\rho_i)$ are the $^{10}$B and $^{11}$B event count for the same bin, respectively. $\Delta N^{^{11}B}(\rho_i)$
can be rewritten using the $^{10}$B fraction in kinetic energy:
\begin{equation}
 \Delta N^{^{11}B}(\rho_i) = \Delta N^{^{11}B}(E^{11}_i) = \frac{1-F_B(E^{11}_i)}{F_B(E^{11}_i)} \Delta N^{^{10}B}(E^{11}_i)\, .
\end{equation}
Here $\Delta N^{^{11}B}(E^{11}_i)$ denotes the $^{11}$B event count in a bin in kinetic energy per nucleon whose limits are obtained
by converting the limits in rigidity of the $i$-th bin to kinetic energy assuming the mass and the charge of $^{11}$B. $E^{11}_i$ is the
kinetic energy per nucleon of a $^{11}$B nucleus of rigidity $\rho_i$. Then, by construction, the first equality in the above
equation follows. The second equality follows from the definition of $F_B(E_i)$ which is the equivalent of eq. \ref{eq:rigfracbins} for 
kinetic energy bins. Note that:
\begin{equation}
  \Delta N^{^{10}B}(\rho_i) \neq \Delta N^{^{10}B}(E^{11}_i)\, ,
\end{equation}
since the limits of the energy and rigidity bins do not correspond for $^{10}$B. Converting the bin limits in energy back to rigidity
but assuming now the mass and the charge of $^{10}$B yields:
\begin{equation}
  \Delta N^{^{10}B}(E^{11}_i) = \Delta N^{^{10}B}(\rho'_i)\, .
\end{equation}
$\rho'_i$ is then the rigidity of a $^{10}$B nucleus having the same kinetic energy per nucleon $E^{11}_i$ of a $^{11}$B
nucleus of rigidity $\rho_i$ (the same relation holds between the limits of the bins centered in $\rho'_i$ and $\rho_i$).
To obtain the explicit relationship between $\rho'_i$ and $\rho$, write $E^{11}_i$ as:
\begin{equation}
  \rho'_i = \frac{A_{10}}{Z} \sqrt{\left(E^{11}_i\right)^2 + 2 m_p E^{11}_i}\, ,
\end{equation}
where $Z$ is the atomic number of boron, $A_{10}$ is the mass number of $^{10}$B and $m_p$ is the proton mass, and then
$E^{11}_i$ as a function of $\rho_i$:
\begin{equation}
  E^{11}_i = \sqrt{\frac{Z^2}{A_{11}^2} \rho_i^2 + m_p^2} - m_p\, ,
\end{equation}
with $A_{11}$ the mass number of $^{11}$B. It follows that:
\begin{equation}
 \rho'_i = \frac{A_{10}}{A_{11}} \rho_i\, .
\end{equation}
The final form of the rigidity-dependent $^{10}$B fraction is then:
\begin{equation}
 F_B(\rho_i) \approx \frac{\Delta N^{^{10}B}(\rho_i)}{\Delta N^{^{10}B}(\rho_i)+\frac{1-\tilde{F_B}}{\tilde{F_B}}\Delta N^{^{10}B}(\rho'_i)}\, ,
 \label{eq:rigfracfinal}
\end{equation}
where the approximated energy-independent value $F_B(E) \approx \tilde{F_B}$ has been used.

Generally speaking, the fraction expressed as a function of rigidity is not constant and depends on the spectral shape.
To account for this a toy Monte Carlo simulation of realistic $^{10}$B and $^{11}$B spectra taken from a galactic 
propagation model has been set up, the resulting event counts have been trimmed to reproduce $\tilde{F_B}=0.35$ and
finally the events have been binned according to their rigidity for both the pure $^{10}$B and pure $^{11}$B hypotheses. 
Knowing the fraction in each rigidity bin of the two binnings one can express the final boron flux in the $i$-th energy
bin as:
\begin{equation}
  \phi^B (E_i) = {F_B}^{10}_i \phi_{10}(E_i) + (1 - {F_B}^{11}_i)\phi_{11}(E_i)\, ,
\end{equation}
where ${F_B}^{10}_i$ and ${F_B}^{11}_i$ are the $^{10}$B fraction obtained from the toy Monte Carlo in the $i$-th rigidity bin for pure $^{10}$B 
and pure $^{11}$B hypotheses respectively, and $\phi_{10}(E_i)$ and $\phi_{11}(E_i)$ are the experimental fluxes for the pure 
$^{10}$B and pure $^{11}$B hypotheses respectively (see Section \ref{sect:boronisotopes}).

To assess the difference between the $^{10}$B fraction as a function of kinetic energy per nucleon and as a function of rigidity,
eq. \ref{eq:rigfracfinal} can be computed at high energies. Above few GeV/n, where the spectrum can be well described with a 
power-law function with index $\gamma$, eq. \ref{eq:rigfracfinal} gives a $^{10}$B fraction 
\begin{equation}
  F_B (\rho_i) \approx \frac{1}{1 + \frac{1-\tilde F_B}{\tilde F_B} (A_{10}/A_{11})^{-\gamma}} \simeq 0.288\, ,
\end{equation}
which is in agreement with the value obtained from the toy Monte Carlo and differs from $\tilde{F_B} = 0.35$ by about 18\%.

\end{document}